# MnEdgeNet—Accurate Decomposition of Mixed Oxidation States for Mn XAS and EELS L2,3 Edges without Reference and Calibration

Huolin L. Xin[*], Mike Hu

Department of Physics and Astronomy, University of California, Irvine, CA 92697, United States

*Correspondence should be addressed to H.L.X. (huolin.xin@uci.edu)

**Abstract**: Accurate decomposition of the mixed Mn oxidation states is highly important for characterizing the electronic structures, charge transfer and redox centers for electronic, electrocatalytic and energy storage materials that contain Mn. Electron energy loss spectroscopy (EELS) and soft X-ray absorption spectroscopy (XAS) measurements of the Mn L2,3 edges are widely used for this purpose. To date, although the measurement of the Mn L2,3 edges are straightforward given the sample is prepared properly, an accurate decomposition of the mix valence states of Mn remains non-trivial. For both EELS and XAS, 2+, 3+, 4+ references spectra need to be taken on the same instrument/beamline and preferably in the same experimental session because the instrumental resolution and the energy axis offset could vary from one session to another. To circumvent this hurdle, in this study, we adopted a deep learning approach and developed a calibration-free and reference-free method to decompose the oxidation state of Mn L2,3 edges for both EELS and XAS. To synthesize physics-informed and ground-truth labeled training datasets, we created a forward model that takes into account plural scattering, instrumentation broadening, noise and energy axis offset. With that, we created a 1.2 million-spectrum database with a three-element oxidation state composition label. The library includes a sufficient variety of data including both EELS and XAS spectra. By training on this large database, our convolutional neural network achieves 85% accuracy on the validation dataset. We tested the model and found it is robust against noise (down to PSNR of 10) and plural scattering (up to $t/\lambda = 1$). We further validated the model against spectral data that were not used in training. In particular, the model shows high accuracy and high sensitivity for the decomposition of Mn3O4, MnO, Mn2O3 and MnO2. In particular, the accurate decomposition of Mn3O4 experimental data shows the model is quantitatively correct and can be deployed for real experimental data. Our model will not only be a valuable tool to researchers and materials scientists but also can assist experienced electron microscopists and synchrotron scientists in automated analysis of Mn L edge data.

# Introduction

X-ray absorption spectroscopy (XAS)[1] and electron energy loss spectroscopy (EELS)[2,3] are two techniques that can probe the unoccupied electronic states providing bonding information of materials. In particular, the L2,3 edges are widely used to determine the oxidation state of transition metals[1,4,5]. The transition metal L2,3 edges probe the unoccupied $d$ orbitals and therefore the edge onset and the edges' fine structures and shapes are sensitive to the oxidation state of the $d$-block metal ions, in particular the 3$d$ transition metals, such as V, Ti, Mn, Fe, and Ni[5-8]. For example, using the near edge fine structures in the Mn L2,3 edges, the oxidation states of Mn ions in a material can be determined by decomposing the spectrum into a linear combination of Mn2+, Mn3+, and Mn4+ reference spectra[9,10]. This decomposition, in principle, is simple but in reality, it is non-trivial because the energy axis is not always calibrated, and the instrument/beamline does not always have the instrumental broadening. Without proper calibration, an energy offset is present between the experimental spectrum and the references which prevents accurate oxidation states decomposition. In order to avoid the problem, standard reference samples such as MnO, Mn2O3, MnO2 need to be measured in the same experimental session to avoid any energy offsets as well as change in instrumental broadening[9,11]. Still with this procedure, there are other factors that could prevent the proper energy axis calibration for example temperature fluctuations would result in an energy shift in the monochromator. Basically, if the XAS measurements are separated multiple hours in time, the spectra taken could have a slight energy offset. In EELS, the energy offset could change more rapidly and is more unpredictable than XAS. Typically, the energy offset is very sensitive to the DC stray field. For example, the passing of a truck or the movement of a nearby elevator, all could change the energy offset if the TEM is not fully shielded. This problem is now mitigated with the dualEELS instruments but there are still many single EELS instruments under active service. Moreover, all historical data were acquired without the dualEELS correction. In addition, nonlinearity of parallel EELS spectrometer is present in EELS in a nontrivial way because the nonlinearity is not only present in the dispersion device, i.e. the magnetic prism. There is another complex nonlinearity present in the magnification lenses, a series of quadrupole. Therefore, it is extremely difficult to calibrate the energy onset of EELS edges unless strict protocols are followed as described by Tan et al[11].

Another complication is that EELS' near-edge fine structures change with sample thickness due to plural scattering. As the sample gets thicker, signals close to the edge onset would be multiple scattered to higher energy losses. This would result in a shape change of the spectrum[11]. For example, for latter $3d$ transition metals's L2,3 edges, as the sample gets thicker, the L2/L3 ratio increases—this problem has rendered the reference-free L2,3 ratio method inaccurate for EELS[11]. In addition, for XAS, the background and the near edge structures could be different between the TEY and TFY modes. That also renders the L2,3 ratio method unreliable. Moreover, for early $3d$ transition metals, there are no established reference-free methods because of the L2,3 anomaly.

For both EELS and XAS spectroscopy, one interesting observation is that human operators with sufficient training can identify spectral features and assign oxidation states to transition metal L2,3 edges with high confidence. This points to the direction that deep learning could be successful in solving the L2,3 oxidation state decomposition problem. Pate et al in 2021 discussed using deep learning to denoise high frame rate spectra.[12] Chatzidakis and Botton in 2019 introduced the idea of translation-invariance for classifying EELS edges.[13] They built a convolutional neural network (CNN) for oxidation state classification and showed that with a translation-invariant training, moving the energy axis does not change the Mn 2+, 3+, 4 oxidation state classification. This is a very important step in demonstrating that spectral features are like spatial features in images—they can be classified by a CNN network regardless of their absolute energy positions in the spectrum.

However, there are still problems remained to be resolved: 1) how to quantitatively decompose mixed oxidation states; 2) is it possible to build one model that work for both XAS and EELS spectroscopy that have drastically different energy resolution; 3) is it possible to build a model that is not affected by plural scattering, i.e. the thickness effect in EELS.

To address the three challenges defined above, in this study, we present a reference-free, calibration-free deep learning approach to determine the accurate oxidation states decomposition of 3$d$ transition metal based on the L2,3 near edge fine structures. To demonstrate the validity of the method, in this study, we use Mn as an example because Mn is technologically important in catalysis, energy storage and electronic materials. Determining the composition of the mixed oxidation states is extremely important for understanding the charge transfer phenomenon happening at the device interfaces. The method we present in this study is not a simple classification of Mn2+, 3+ and 4+ edges but an accurate and quantitative decomposition of the mixed Mn oxidation states. Instead of having a classification/binary type label, we created a three-element ground truth vector that quantitatively describe the composition of Mn2+, 3+ and 4+ in each Mn spectrum, i.e. [%$Mn_{2+}$, %$Mn_{3+}$, %$Mn_{4+}$].

To achieve this goal, we built a 1.2 million-spectrum ground truth labeled library with 50% XAS data and 50% EELS data. In building the mixed oxidation state library, we paid special attention to normalizing the Mn L2,3 edges correctly, and including experimental-like uncertainties such as both Gaussian and Lorentzian type instrumental broadening, energy offset and detector noise. To include the plural scattering effect in the training library, we developed a forward model to correctly introduce the thickness effect to the L2,3 edges. Using this physics-informed large training library, we show that the deep convolutional regression model we trained is robust against plural scattering and noise. The overall accuracy of model in determining the mixed valence state reaches 85% on the validation data set. We also validated the data on "unknown unknowns", i.e. Mn3O4 spectra that have never been used for training—the accurate decomposition of Mn3O4 experimental data shows the model is quantitatively correct and can be deployed for real experimental data.

## Methods

In this method section, we will describe 1) how to build a ground-truth oxidation state labeled Mn edge library, 2) how to construct the neural network, and 3) how to train it.

For building the library, the technical challenges lie in 1) how to obtain a wide variety of XAS and EELS Mn2+, Mn3+, Mn4+ reference spectra; 2) how to normalize or ratio the 2+, 3+ and 4+ spectra correctly; 3) how to include the EELS's plural scattering effect (thickness effect) into the training sets; 4) how to include the various experimental uncertainties including instrumental broadening, energy offset, detector noise, etc. In the following subsections, we will address each aforementioned challenge.

**Collection of Mn reference spectra**

To have sufficient varieties of data that can capture the features of the EELS and XAS Mn 2+, 3+ and 4+ edges, in this study, we digitized a large number of experimental EELS and XAS Mn spectra that were documented in the literature. In Figure 1, we presented all spectra that were used for making the training library. (The Mn 2.67+ spectrum was not included in the training library.) In Table 1, we listed the compounds for which we digitized the spectra and their original references.

All data are standardized to range from 630.5 eV to 669.4 eV with 0.1 eV increments (338 data points). For missing data, the left side of the spectra is padded with zero and the right side is padded with the end value of the spectra. Fig xx shows examples of the standardized data.

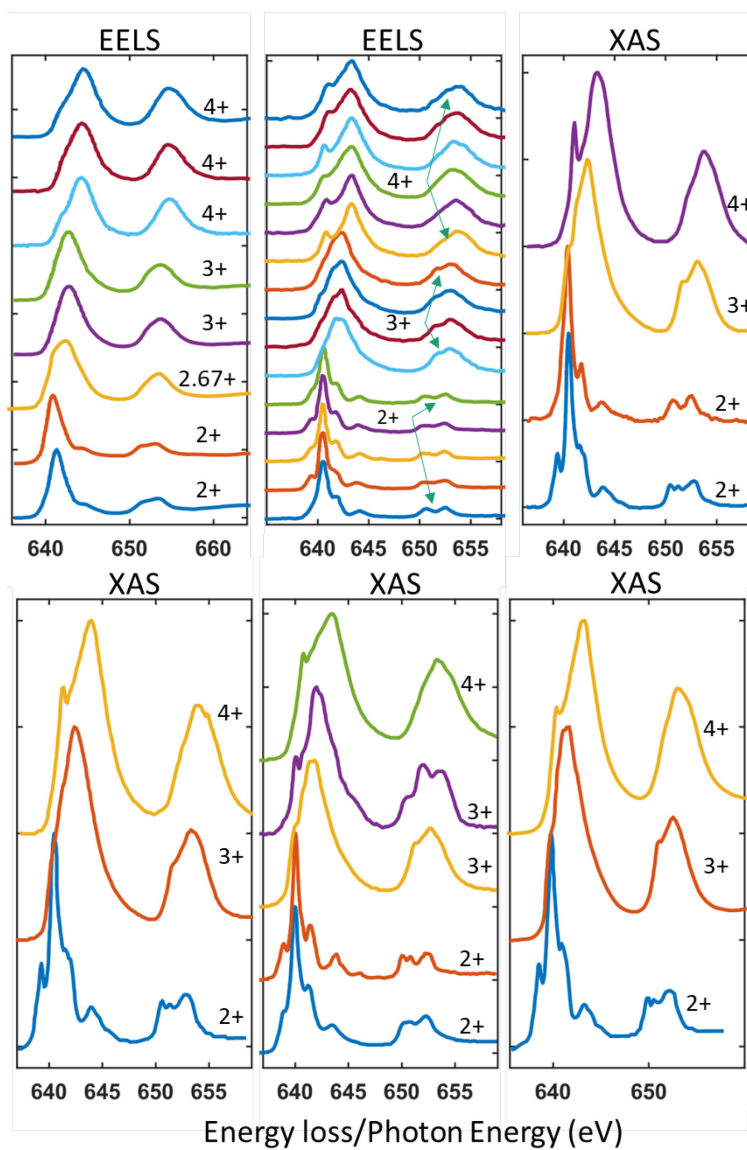

Figure 1. The presentation of the EELS and XAS Mn L2,3 edges included in making the training library. The Mn 2.67+ presented is not included in the training library.

Table 1. The compound information and references of the Mn L2,3 edges.

| Oxidation state | Compounds | References |
|---|---|---|
| 2+ | MnO (Manganosite), MnV2O4, (LiMnPO4)Lithiophilite, (MnSiO$_3$) Rhodonite, MnF2, (MnCO3) Rhodochrosite, YBaMn3AlO7, | EELS Refs: [11,14] |
| 3+ | MnOOH(Manganite), Mn2O3, ((Mn,Fe)$_2$O$_3$) Bixbyite, (Ca$_4$Mn$^{3+}_{2-3}$(BO$_3$)$_3$(CO$_3$)(O,OH)$_3$) Gaudefroyite, LaMnO3 | XAS Refs: [15-18] |
| 4+ | SrMnO3, CaMnO3, MnO2, Todorokite, ((Ni,Co)$_{2-x}$Mn$^{4+}$(O,OH)$_4$ · nH$_2$O) Asbolan, (ZnMn$^{4+}_3$O$_7$ · 3H$_2$O)Chalcophanite, (Mn$^{4+}$O$_2$) Ramsdellite, (Mn$^{4+}$O$_2$) Pyrolusite | |
| 2.67+ | Mn3O4 (note used for training) | |

**Normalization of 2+, 3+, 4+ reference spectra**

In order to quantitatively combine the 2+, 3+, and 4+ Mn spectra, they need to be normalized them to the correct ratio. To achieve that, we normalize the Mn L3 edge according to the *d*-hole number. Elemental Mn has an electron configuration of [Ar] 3*d*5 4*s*2. Therefore, Mn2+, 3+, 4+ have an electron configuration of [Ar] 3*d*5, [Ar] 3*d*4, [Ar] 3*d*3. Because *d* shell can hold 10 electrons, the number of *d* holes for Mn 2+, 3+ and 4+ are 5, 6, 7 respectively. Therefore, the area under the L3 peak and above the continuous background shall be proportional to the *d* hole number. The continuous background under the L2,3 edge can be modeled by two step functions with a step height that follows the 1:2 population ratio. (The filled 2*p*3/2, 2*p*1/2 orbitals have a population ratio of 1:2). The *d* hole area can be calculated after the background is subtracted from the spectrum (Figure 2). With this procedure to find the *d*-hole area, we can correctly ratio the 2+, 3+ and 4+ spectra.

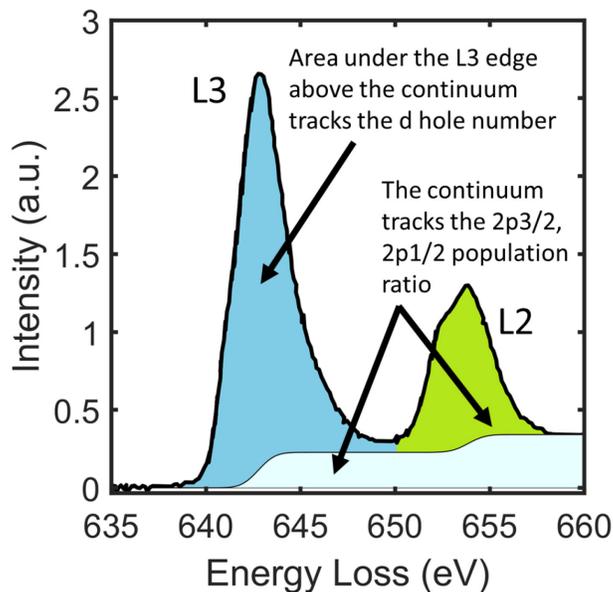

Figure 2. Schematics showing how to extract the *d*-hole area under the L3 edge.

**Ground truth labeled library**

After the *d*-hole ratio normalization, we can correctly combine the Mn 2+, 3+ and 4+ component spectra to form a new spectrum with the known ground truth oxidation state composition and oxidation state as the following

$$s = xMn^{2+} + yMn^{3+} + zMn^{4+}$$
$$where\ x + y + z = 1$$
ground truth oxidation state composition = [*x*, *y*, *z*]
ground truth average oxidation state = 2*x* + 3*y* + 4*z*.

In making the ground truth labeled spectra, we only combined Mn spectral components that were digitalized from the same publication source. The reason being that the spectra to be combined shall share the same instrumental resolution.

The composition of the training library is detailed in Table 2. A total of 1,210,000 synthetic spectra are included in the library.

Table 2. The composition of the ground truth labeled library.

| Components | Occurrence | Type |
| --- | --- | --- |
| Single component 2+, 3+, or 4+ | 2.5% | 50% EELS and 50% XAS |
| Two components (2+, 3+), (2+, 4+) or (3+, 4+) | 48.75% | 50% EELS and 50% XAS |
| Three components (2+, 3+, 4+) | 48.75% | 50% EELS and 50% XAS |

**Instrumental broadening**

The instrumental broadening of EELS spectra includes two major contributions. The first contribution primarily comes from the thermal broadening at the electron source and the first cross over due to the space-charge effect. This type of broadening is typically characterized by a Gaussian type broadening function. The second contribution happens at the detector. The light diffusion in the sinterlator and the optical coupler introduce a long-tail broadening effect which can be characterized by a Lorentzian function. For XAS, similar short range and long-range broadening happens due to the monochromator. Therefore, we introduce a two-parameter controlled instrument broadening kernel aka the point spread function, PSF(E) as the following

$$\text{PSF}(E, w, \sigma) = L(E,w) \otimes G(E,\sigma)$$

where $\otimes$ stands for convolution

$$L(E, w) = \frac{1}{\pi} \frac{1/2w}{E^2 + (1/2w)^2}$$

and

$$G(E, \sigma) = \frac{1}{\sigma\sqrt{2\pi}} e^{-\frac{1}{2}\left(\frac{E}{\sigma\sqrt{2}}\right)}$$

Basically, the instrumental point spread function is a convolution of a Gaussian function with a Lorentzian function. The full width at maximum (FWHM) of the Lorentzian function is $w$ and the FWHM of the Gaussian function is $2\sqrt{2ln(2)}\sigma$. The combined FWHM is equal to $\sqrt{FWHM_{Lorentzian} + FWHM_{Gaussian}}$. It is worth noting that the inclusion of the Lorentzian tails in the point spread kernel is very important for making the synthesized spectra resemble the experimental ones. An example of such broadening effect on a Mn2+ L2,3 edge is shown in Figure 3.

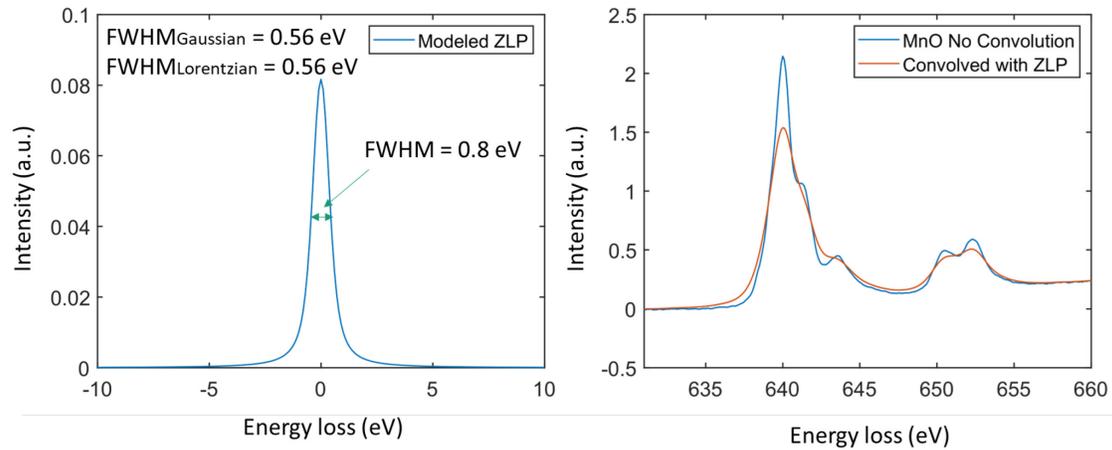

Figure 3. An example of instrumental broadening effect on the Mn L2,3 edge.

## Plural scattering in EELS

If the single scattering probability function is $P(E)$, plural scattering as a function of thickness, $t$, in EELS can be described by the following differential equation

$$\frac{dS(E,t)}{dt} = \int S(E')P(E-E')dE'$$

and the boundary condition is $S(E, t=0) = \delta(E)$ in the ideally monochromatic condition. In the practical situation where the incoming electron has an energy spread, we can use the point spread function given in the last section as the initial energy profile, i.e.

$$S(E, t=0) = PSE(E, w, \sigma)$$

Once we obtain a numerical representation of $P(E)$, the spectral function, $S(E, t)$ at any given thickness, $t$ can be numerically calculated.

Using this equation, it allows us to calculate the low-loss spectral function numerically. Once we obtain the low-loss spectral function, the core-loss spectrum is a convolution of the core-loss single scattering probably distribution, $P_{core-loss}(E)$, with the low-loss spectral function, i.e. $S(E, t)$.

Figure 4 shows the change of the low-loss function as function of normalized thickness ($t/\lambda$, $\lambda$ is the inelastic mean free path) and how the Mn L2,3 edge evolves.

In this modeling, we use an average plasmon loss energy of 25 eV and approximate the $P(E)$ by an asymmetric function where the left side is a Gaussian function, and the right size is a Lorentzian loss function. To be more exact, we also modeled the Mn M edge and superimposed it onto the plasmon loss.

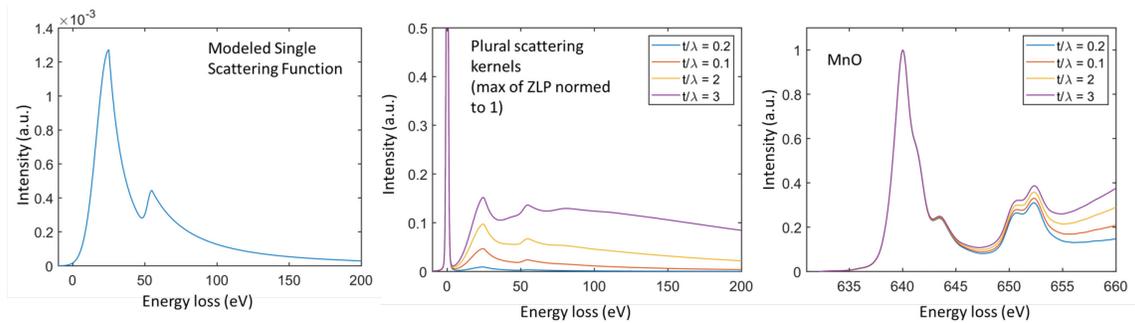

Figure 4. The modeling of the plural scattering for Mn containing compound and its effect on the spectral shape of Mn L2,3 edges.

## Other augmentations: energy shift and noise

Both EELS and XAS are subject to the issues of inaccurate energy axis. To take this into account, we apply a random shift augmentation of the energy axis for the ground truth labeled spectra. With this augmentation, the model become translation invariant—it is only sensitive to the spectral shape and it is insensitive to the absolute energy onset of the L2,3 edge.

For noise, we have modeled the noise as a white noise (Gaussian noise) with a salt and pepper noise (impulse noise). Both noises are additive to the spectrum. We use the linear definition of PSNR as:

$$PSNR = \frac{Max\ Signal}{\sqrt{Mean\ Sequare\ Error}}$$

**Summary of augmentation**

In the table below, we summarize the augmentation operations done to the ground truth labeled library.

Table 3. A summary of the augmentation operations and occurances.

| Type of augmentation | Probability of application | Parameters |
|---|---|---|
| Instrumentation broadening | 80% | Gaussian: FWHM uniformly distributed between 0.01 and 1.5 eV<br>Lorentzian: FWHM uniformly distributed between 0.1 and 0.4 eV |
| Plural scattering | 80% | Normalized thickness $t/\lambda$ uniformly distributed between 0 and 1 |
| Shifts | 100% | Uniformly distributed between -4 eV and +4 eV |
| Noise | 50% | PSNR ranges from 10 to 30 |

**Network structure**

How our brains process or identify a spectral feature is very similar to recognizing spatial features in an image. Inspired by this, we adopted the convolution layers that are used in image classification for feature extraction. Then we connected the features with a fully connected layer (also known as dense layers) for composition regression. The input is the one-dimensional spectrum, and the output is a 3-element composition vector (Figure 5). We call this network a convolutional regression net (CRN). Different from a classification network, a regression network's outputs are continuous numbers rather than binary number. Therefore, we used the mean square error function as the loss function.

For feature extractions, we use three convolutional layers followed by leaky relu and maxpooling. The final layer outputs 41*128=5248 filtered features. In the regression layers, we used three fully connected layers with 2048, 512 and then 3 neurons with leaky relu in between. The final output is a softmax normalization of the final 3-neuron layer to ensure that sum of the composition vector is equal to 1.

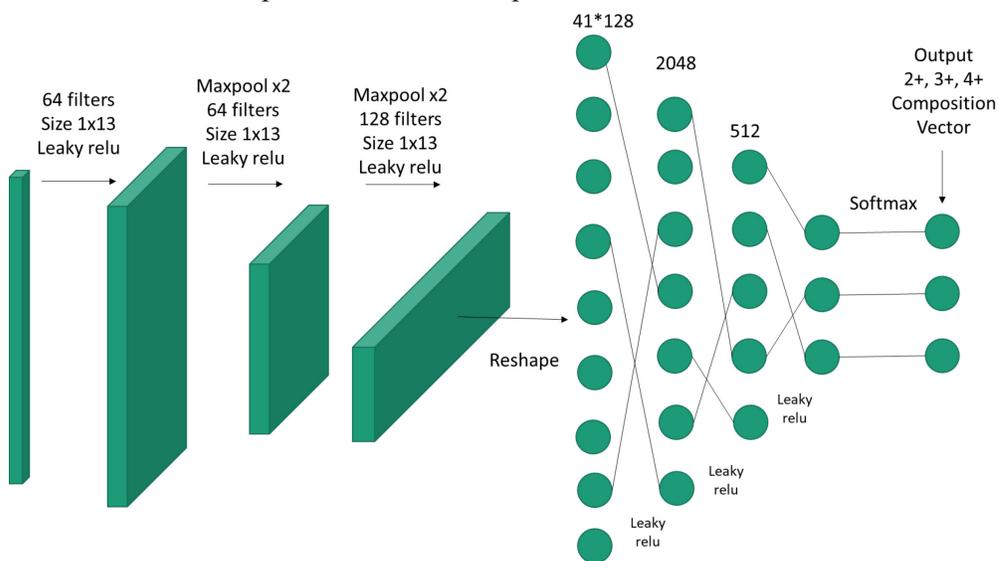

Figure 5. The structure of the convolutional regression net for mixed oxidation state decomposition.

### Training

All spectra are subtracted by the mean and divided by the standard deviation before entering the network. Dropout are added to each layer before maxpooling with a dropout rate of 0.1. Adam, an algorithm for first-order gradient-based optimization of stochastic objective functions, based on adaptive estimates of lower-order moments was used for learning. The learning rate is set at 8E-5. Batch size is 32. As shown in Fig. xx, the model converges quickly; therefore, only 3 epochs of training was done to avoid overfitting.

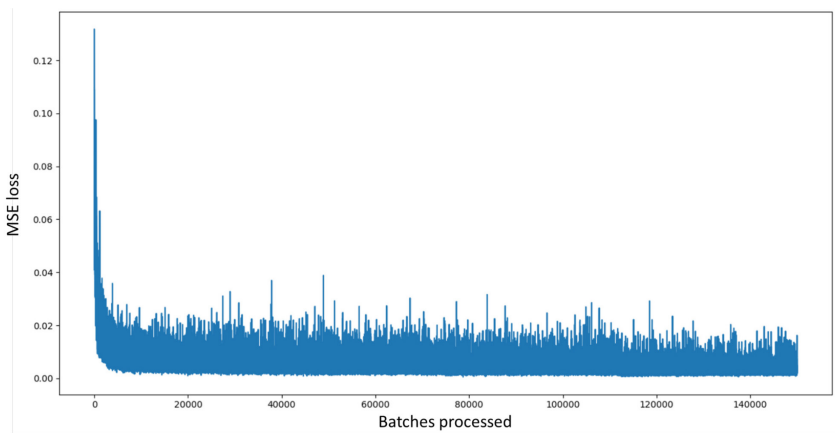

Figure 6. The MSE loss as a function of batches processed.

### Validation

We performed a 20/80 split of the ground truth labeled library into training and validation data sets. The accuracy of the model is evaluated on the validation set. An accurate prediction is defined as that the predicted oxidation state falls in ±0.1 of the ground truth oxidation state.

We also evaluated our model on known knowns, unknown knowns and unknown unknowns data sets. Known knowns are the data used to build the training data. The unknown knowns are the data with various augmentations such as noise and plural scattering. The unknown unknowns are new experimental data that were never used for training.

## Result & Discussion

**Performance of the model on the validation set.** On the validation set, the trained model reports an accuracy of 85%. Figure 7 shows the scatter plot of the prediction versus the ground truth (2000 spectra were randomly selected from the validation set). The result shows that the model performs reasonably well on the validation data.

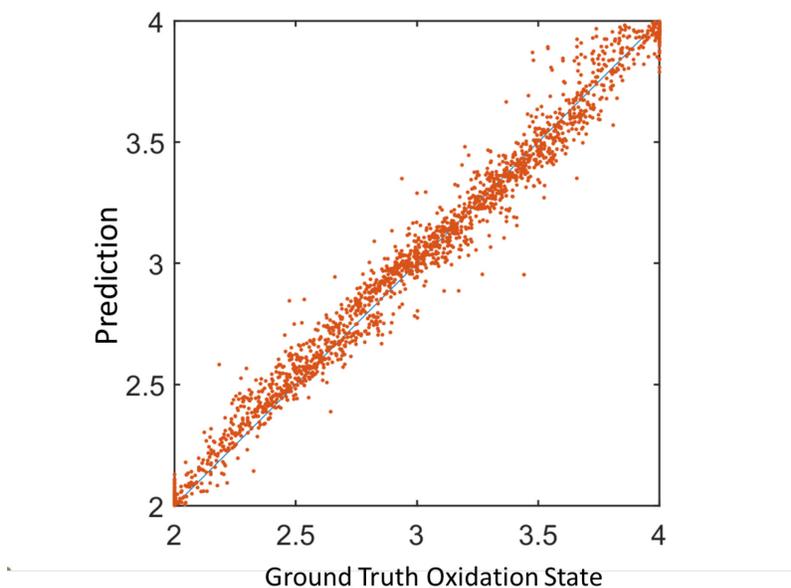

Figure 7. The scatter plot of the model's predicted average oxidation state versus the ground truth.

To more closely look at the performance of the predicted decomposition, we provide a table of the predicted composition as shown in Table 4. It shows that the decomposition is reasonably accurate on the validation dataset.

Table 4. CRN's decomposition performance on validation spectra

| Compound | Predicted [2+,3+,4+] decomposition (%) | Predicted oxidation state |
|---|---|---|
| Mn2O3 | 0.98, 98.5, 0.52 | 3.0 |
| MnF2 | 99.69, 0.31, 0.0 | 2.0 |
| MnF3 | 0.66, 98.87, 0.48 | 3.0 |
| MnO | 98.05, 1.85, 0.1 | 2.02 |
| MnO2 | 0.45, 4.67, 94.88 | 3.94 |
| MnO | 99.6, 0.37, 0.03 | 2.0 |
| MnV2O4 | 99.42, 0.21, 0.37 | 2.01 |
| MnOOH | 0.18, 98.03, 1.79 | 3.02 |
| CaMnO3 | 0.0, 2.4, 97.6 | 3.98 |
| MnO2 | 0.0, 0.01, 99.98 | 4.0 |
| SrMnO3 | 0.0, 0.06, 99.94 | 4.0 |
| 75% Mn3+OOH +25%Mn4+O2 | 0.09, 73.45, 26.46 | 3.26 |
| 75% Mn3+OOH + 25% SrMn4+O3 | 0.25, 70.55, 29.2 | 3.29 |
| 60% Mn4+O2 + 40% SrMn4+O3 | 0.0, 0.02, 99.98 | 4.0 |

**Plural scattering.** To show how the model performs with the interference of plural scattering, we tested the thickness effect on MnO, Mn2O3, and MnO2. As shown in Table 5, the results are accurate up to 1.5 inelastic mean free path ($\lambda$) which is larger than maximum augmentation range used in the training dataset.

Table 5. Testing of CRN's decomposition robustness against plural scattering.

| Thickness | Predicted decomposition (%) | Predicted oxidation state |
|---|---|---|
| MnO | | |
| 0 | [99.6, 0.37, 0.03] | 2.0 |
| 0.1 | [99.49, 0.5, 0.01] | 2.01 |
| 0.3 | [99.41, 0.58, 0.01] | 2.01 |
| 0.5 | [99.28, 0.71, 0.01] | 2.01 |
| 0.8 | [98.94, 1.04, 0.02] | 2.01 |
| 1 | [98.64, 1.33, 0.03] | 2.01 |
| 1.5 | [97.28, 2.57, 0.16] | 2.03 |
| 2 | [94.27, 4.14, 1.59] | 2.07 |
| 2.5 | [88.33, 4.42, 7.25] | 2.19 |
| 3 | [80.74, 3.14, 16.12] | 2.35 |
| Mn2O3 | | |
| 0 | [0.66, 99.13, 0.21] | 3.0 |
| 0.1 | [0.57, 99.21, 0.22] | 3.0 |
| 0.3 | [0.6, 99.19, 0.21] | 3.0 |
| 0.5 | [0.69, 99.07, 0.24] | 3.0 |
| 0.8 | [1.25, 98.46, 0.28] | 2.99 |
| 1 | [2.2, 97.54, 0.26] | 2.98 |
| 1.5 | [12.87, 86.66, 0.48] | 2.88 |
| 2 | [43.15, 55.66, 1.19] | 2.58 |
| 2.5 | [52.78, 44.89, 2.33] | 2.5 |
| 3 | [59.93, 36.5, 3.57] | 2.44 |
| MnO2 | | |
| 0 | [0.0, 0.01, 99.98] | 4.0 |
| 0.1 | [0.0, 0.03, 99.97] | 4.0 |
| 0.3 | [0.01, 0.04, 99.95] | 4.0 |
| 0.5 | [0.01, 0.06, 99.93] | 4.0 |
| 0.8 | [0.01, 0.11, 99.88] | 4.0 |
| 1 | [0.02, 0.21, 99.77] | 4.0 |
| 1.5 | [0.11, 1.75, 98.14] | 3.98 |
| 2 | [0.57, 10.88, 88.55] | 3.88 |
| 2.5 | [3.33, 23.44, 73.23] | 3.7 |
| 3 | [16.82, 20.4, 62.78] | 3.46 |

**Performance with noise.** To show how the model performs with the interference of noise, we tested the effect on MnO, Mn2O3, and MnO2. As shown in Table 6, the model is robust down to PSNR = 20. At PSNR = 10, 2+ and 4+ are more stable than 3+.

Table 6. Testing of CRN's decomposition robustness against noise.

| PSNR | Predicted decomposition (%) | Predicted oxidation state |
|---|---|---|
| MnO | | |
| None | [99.6, 0.37, 0.03] | 2.0 |
| PSNR = 30 | [99.97, 0.02, 0.01] | 2.0 |
| PSNR = 20 | [99.99, 0.0, 0.0] | 2.0 |
| PSNR = 10 | [99.68, 0.3, 0.03] | 2.0 |
| PSNR = 5 | [99.74, 0.26, 0.0] | 2.0 |
| PSNR = 3 | [99.95, 0.05, 0.0] | 2.0 |
| Mn2O3 | | |
| None | [0.25, 99.59, 0.16] | 3.0 |
| PSNR = 30 | [0.94, 98.08, 0.98] | 3.0 |
| PSNR = 20 | [1.58, 98.41, 0.01] | 2.98 |
| PSNR = 10 | [22.09, 77.91, 0.0] | 2.78 |
| PSNR = 5 | [89.7, 9.85, 0.44] | 2.11 |
| PSNR = 3 | [0.0, 0.0, 99.99] | 4.0 |
| MnO2 | | |
| None | [0.0, 0.01, 99.98] | |
| PSNR = 30 | [0.0, 0.05, 99.95] | 4.0 |
| PSNR = 20 | [0.0, 0.0, 100.0] | 4.0 |
| PSNR = 10 | [0.09, 10.54, 89.37] | 4.0 |
| PSNR = 5 | [4.73, 0.13, 95.14] | 3.89 |
| PSNR = 3 | [0.0, 0.0, 100.0] | 3.9 |

**Validation of the model on experimental data that were not used for training.**

Validation on unknown unknowns is critical for understanding the accuracy and robustness of a machine learning model.

**Testing on Mn3O4.** One of the compounds, for which we have experimental data on, but was not used for training was Mn3O4. It has a mixed oxidation state of Mn2+ and Mn3+

with a theoretical ratio of 1:2. It gives an average oxidation state of +2.67. Figure 8 shows the predicted oxidation state as a function of thickness and the oxidation state decomposition is shown in Table 7. The model predicts the correct ratio between 2+/3+ with a small error. The prediction starts to deviate from the ground truth at $t/\lambda =1.5$ which is larger than the maximal augmentation used for training. Therefore, the reduced performance is expected.

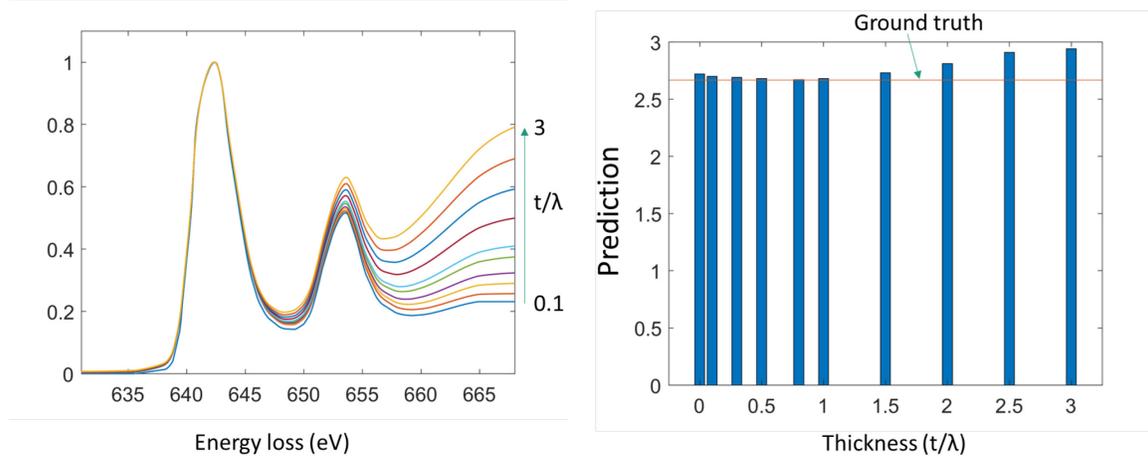

Figure 8. Validation on Mn3O4 as a function of thickness.

Table 7. Testing of CRN's decomposition robustness against plural scattering on unknown unkowns (Mn3O4).

| Thickness | Predicted decomposition (%) | Predicted oxidation state |
|---|---|---|
| 0.0 | 32.03, 64.39, 3.59 | 2.72 |
| 0.1 | 32.00, 64.02, 2.99 | 2.7 |
| 0.3 | 33.81, 63.79, 2.4 | 2.69 |
| 0.5 | 34.31, 63.67, 2.02 | 2.68 |
| 0.8 | 34.39, 64.03, 1.59 | 2.67 |
| 1.0 | 34.16, 64.1, 1.74 | 2.68 |
| 1.5 | 32.28, 64.1, 1.74 | 2.73 |
| 2.0 | 30.24, 58.1, 11.67 | 2.81 |
| 2.5 | 27.73, 53,69, 18.57 | 2.91 |
| 3.0 | 28.38, 49.6, 22.02 | 2.94 |

**Further testing on the influence of noise.** The ratio between 2+ and 3+ stays close to 1:2 for PSNR down to 10. When below 10, the composition ratio starts to deviate from the theoretical ground truth as expected (the noise augmentation range is PSNR = [10,30]).

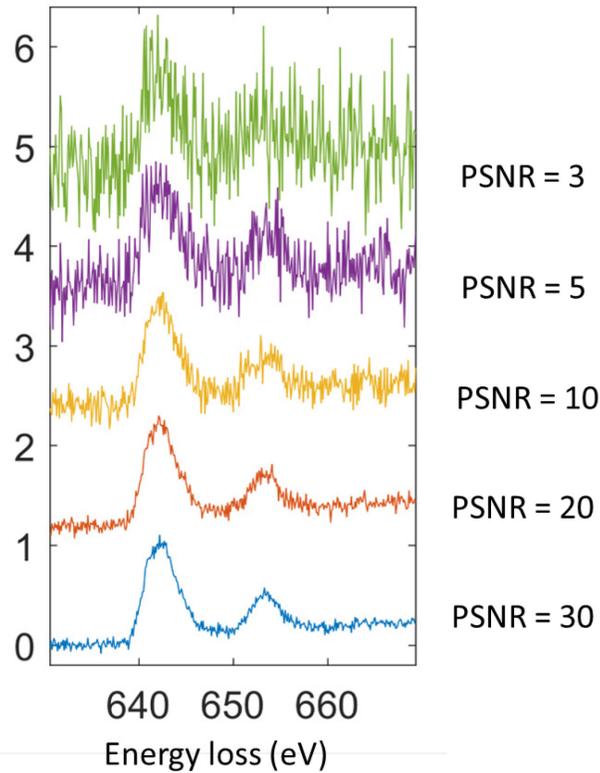

Figure 9. The unkown unkown Mn3O4 spectra as a function of noise.

Table 8. Testing of CRN's decomposition robustness against noise on unknown unkowns (Mn3O4).

| PSNR | Predicted decomposition (%) | Predicted oxidation state |
|---|---|---|
| 30 | 29.41, 67.05, 3.54 | 2.74 |
| 20 | 36.23, 63.76, 0.01 | 2.64 |
| 10 | 38.01, 61.75, 0.24 | 2.62 |
| 5 | 47.8, 52.09, 0.1 | 2.52 |
| 3 | 66.91, 21.65, 11.45 | 2.45 |

**Sensitivity and accuracy validation on Mn3O4 with vacancies on the tetrahedral sites.** We tested the accuracy and sensitivity of our model using two Mn3O4 EELS spectra documented in ref. [9]. The two spectra are shown in Figure 10. The small different in the L3 edge indicates that the nanosized Mn3O4 has some slightly more $Mn^{2+}$ and less $Mn^{3+}$ than the bulk Mn3O4. The documented ratio of $Mn^{3+}/Mn^{2+}$ is 2 for the bulk sample and 1.6 for the nanosized sample in ref. [9]. Our model accurately captures this change . For the nanosized sample, our model's predicted decomposition clearly shows the reduction of $Mn^{3+}$ composition and increase of $Mn^{2+}$ composition. The ratio of $Mn^{3+}/Mn^{2+}$ is predicted to be 1.59 which is almost the same as the documented value. This test shows that our model is accurate and sensitive to small changes in the spectrum.

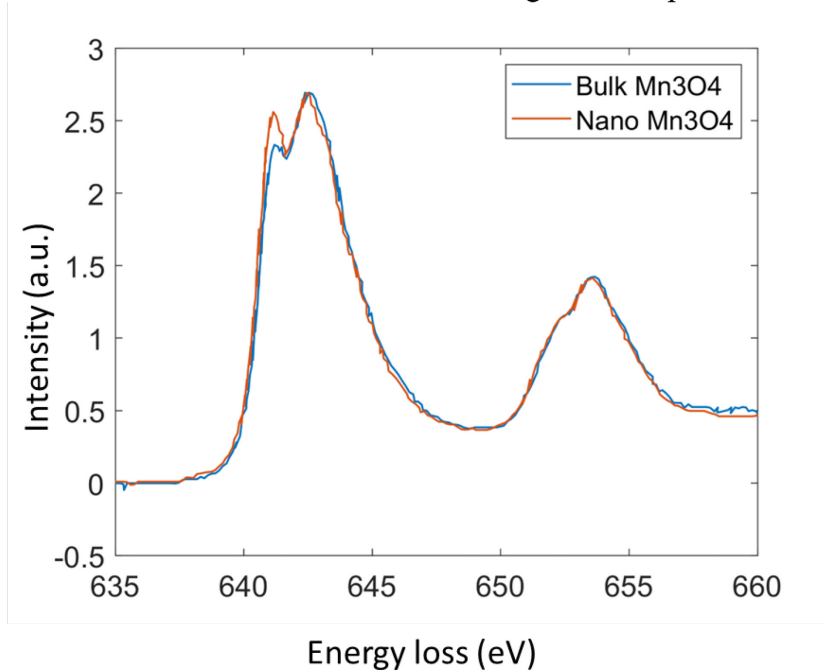

Figure 10. The EELS Mn L2,3 edges of bulk Mn3O4 vs naoszied Mn3O4.

Table 9. Testing of CRN's decomposition sensitivity on unknown unkowns (Mn3O4).

| PSNR | Predicted decomposition (%) | Ratio documented in ref. [9] |
|---|---|---|
| Bulk | 33.92, 64.33, 1.75 | 0.33, 0.67, 0 |
| Nano | 38.1, 60.51, 1.39 | 0.4, 0.6, 0 |

**More on XAS and EELS data not used in training**.

To further test the model on unknown unknowns. We collected more EELS and XAS data from literature and experiments with very different energy resolutions. All data shown in Figure 11 were not used for training. Figure 11 shows the oxidation decomposition of EELS/XAS Mn L2,3 edge inferred by our model. All predictions are within reasonable errors of the ground truth. It is worth noting that the model is effective on both XAS and EELS spectra. The XAS and EELS have very different energy resolutions. Within the XAS, the TEY and PFY also have noticeable differences in the fine structures. In addition, the energy onsets are all different. However, as shown, our model remains translation invariant and is robust enough to correctly decompose their oxidation states.

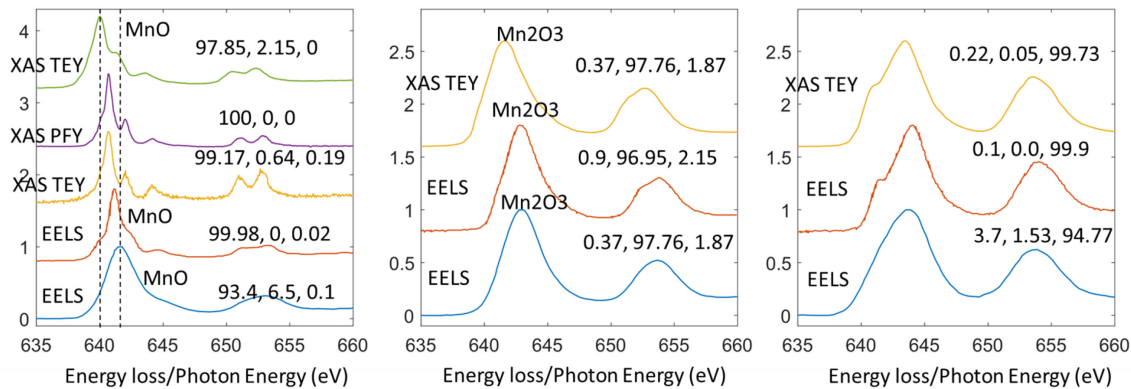

Figure 11. Validation CRN's decomposition sensitivity on data not used for training.

## Conclusion

Determination of the oxidation states of Mn is highly important for characterizing the electronic structures and redox centers for electronic, electrocatalytic and energy storage materials. EELS and XAS measurement of the Mn L2,3 edges are widely used for this purpose. To date, although the measurement of the Mn L2,3 edges are straightforward given the sample is prepared properly, an accurate decomposition of the mix valence states of Mn remains non-trivial. For both EELS and XAS, 2+, 3+, 4+ references spectra need to be taken on the same instrument/beamline and preferably in the same experimental session because the instrumental resolution and the energy axis offset could vary from one session to another. To circumvent this hurdle, in this study, we adopted a deep learning approach and developed a calibration-free and reference-free method to decompose the oxidation state of Mn L2,3 edges for both EELS and XAS. To synthesize physics included and ground-truth labeled training data, we created a forward model that takes into account plural scattering, instrumentation broadening, noise and energy axis offset. With that we created a 1.2 million-spectrum database with oxidation states labeled. The library includes sufficient variety of data including both EELS and XAS spectra. By training on this large database, our convolutional neural network achieves 85% accuracy on the validation dataset. We tested the model and found it is robust against noise (down to PSNR of 10) and plural scattering (up to $t/\lambda = 1$). We further validated the model against spectral data that were not used in training. In particular, the model shows high accuracy and high sensitivity for the decomposition of Mn3O4, MnO, Mn2O3 and MnO2. In particular, the accurate decomposition of Mn3O4 experimental data shows the model is quantitatively correct and can be deployed for real experimental data. Our model will not only be a valuable tool to researchers and materials scientists but also can assist experienced electron microscopists and synchrotron scientist in automated analysis of Mn L edge data.

# Acknowledgement

This work is supported by the Office of Basic Energy Sciences of the U.S. Department of Energy, under award no. DE-SC0021204.

# Author Contributions

HLX conceived the idea. All authors designed and carried out the experiments. HLX wrote the manuscript.

# Competing interests

The authors declare no competing interests.

# Data availability

Data and code are available upon request.